# Coupling Analysis of the Surface States in Periodic Microwave Transmission Line

**L. Ivzhenko** [1,2], A. Girich [2], M. Baranowski [3], A. Kharchenko [2], S. Mieszczak [1], S. Polevoy [2], S. Tarapov [2,4], M. Krawczyk [1], and J. Kłos [1]

[1] Adam Mickiewicz University, Institute of Spintronics and Quantum Information,
Faculty of Physics, Uniwersytetu Poznańskiego 2, 61-614 Poznań, Poland

[2] O. Ya. Usikov Institute for Radiophysics and Electronics NAS of Ukraine,
Ac. Proskura str. 12, 61085, Kharkiv, Ukraine

[3] Adam Mickiewicz University, Faculty of Physics, Uniwersytetu Poznańskiego 2, 61-614 Poznań, Poland

[4] V. N. Karazin Kharkiv National University, 4 Svobody Sq., 61022, Kharkiv, Ukraine

e-mail: ivzhenko@amu.edu.pl

*Abstract* – The spectrum of electromagnetic waves in periodic linear structures, such as periodic waveguides or chains of microelements i.e. spheres, cavities, exhibit the sequence of stop bands for propagating waves. Breaking the translational symmetry of the periodic microstrip can also lead to the localization of the microwaves at the microstrip edge. In this paper, we investigated periodic microstrip transmission line represented as 1D photonic crystal operating at the GHz frequencies. On the ground of topology, we explain the condition of surface state existence. The transmission measurements and numerical calculations support our theoretical predictions. Moreover, we show that in the symmetric microstrip the surface states split into symmetric and antisymmetric modes due to evanescent wave coupling between the modes localized on the opposite sides of the microstrip. Interestingly, both modes offer significant microwave transmission inside the frequency gap, which is promising for applications.

## I. INTRODUCTION

The surfaces/edges of periodic structures are a natural defect that breaks a transnational symmetry of the crystal and can be suitable for wave localization. The wave eigenmodes localized on these peculiar defects are called Bloch *surface states* and can be observed for different kind of periodic structures (e.g. for photonic crystals (PC) [1 - 3], electronic superlattices [4] or magnonic crystals [5]). Their frequencies lay in the ranges of frequency gaps, making them distinguishable in the frequency domain from the bands.

The microstrip transmission line (MSTL) composed of two alternatively changing sections of different width can be considered as a prototypical 1D photonic crystal [6]. The fabrication and experimental studies of microwave systems are definitely simpler then photonic systems and provide the same fundamental knowledge about the nature of electromagnetic (EM) waves in periodic system. The bulk parameters determining the dispersion relation of EM waves for microstrip PC are the length (thickness in 1D PC) of each section (layer) and their wave impedances, dependent on the width of each section (the electric permittivity of each layer). For truncated structures of given bulk properties we can define the so called *surface impedance*, which can be used for determination condition of existence of surface or interface states at the microstrip edge or at the junction of two semi-infinite microstrips PCs, respectively.

In this paper we investigate experimentally and numerically the conditions of appearance of edge modes in periodic microstrip. Using the arguments known in photonics, we explain the properties of edge observed in MSTL. We investigate the transmission spectra of the structure paying particular attention to the EM wave tunneling through the surface states located in the frequency gaps. We discuss the conditions of existence of microwave surface states in microstrip, taking into account topological characteristics of the band structure. The particular attention is paid on effect of evanescent coupling for edge modes. For the MSTL symmetric with respect to the flipping of both endings, the edge modes appear in pairs, at very close frequencies. One can consider the MSTL as a band-pass filter for a frequency-division multiplexing, which additionally allows to introduce the reference pilot signal as a edge mode tunnelled in stop band.

## II. STRUCTURE UNDER STUDY AND NUMERICAL CALCULATIONS

The investigated periodic microstrip transmission line (MSTL) consists of 5 connected cells (Fig. 1(a)) and bounded by two edge cells with other parameters. The parameters of the system were chosen in such a way as to ensure the existence of two surface states in the band gap of MSTL. Each internal cell has a length of $d = 16$ mm and consists of a section of a wider microstrip (width of $w_1 = 4$ mm) with a length of $l = 10$ mm and a narrower one with a width of $w_2 = 1.12$ mm (50 Ohm impedance). The edge cells have a length of $d_0 = 11.5$ mm, and their central wide section (width of $w_0 = 9$ mm) has a length of $l_0 = 1.5$ mm, with the same narrow section. The entire structure is located on a substrate Neltec of thickness $u = 0.381$ mm with permittivity $\varepsilon = 2.2 + j\,0.0012$. The metal microstrip elements and the ground plane are 0.35 μm thick.

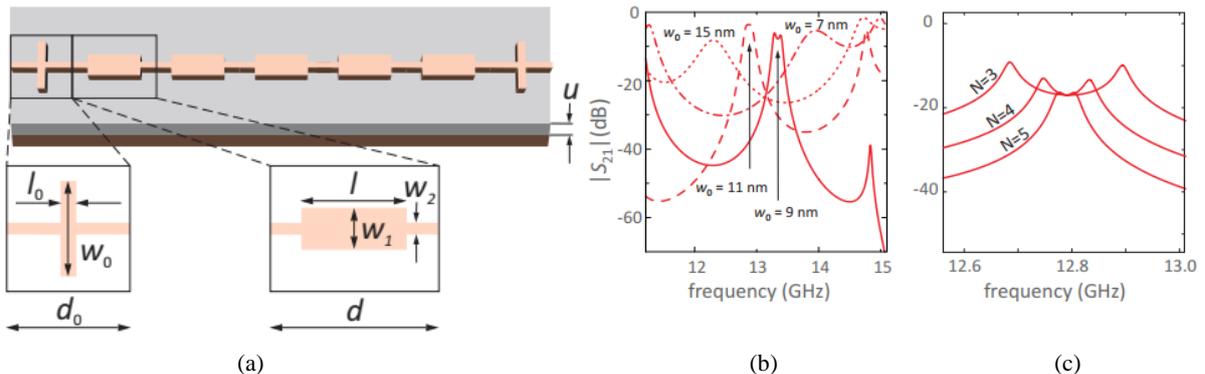

(a) (b) (c)

Fig. 1. Periodic MSTL under study. (a) The geometry of the system; (b) the numerically simulated transmission coefficient spectrum on central section width $w_0$ of edge cells (calculated by MEEP) and (c) the splitting of the S and A type of the surface modes on the number of MSTL cells $N$ (lumped-element calculations).

In order to model the influence of the edge cell parameters on the surface state the transmission spectra of the MSTL structure were calculated numerically. The spectra were calculated by the open-source software package MEEP and by the semi-analytical model based on lumped-element. We consider a cascade of symmetrical two-terminal networks representing the MSTL cells. Each network consists of a shunt capacitance $C$, input and output phase shifters. The shunt capacitance is related to the area of one microstrip above the ground plane, while the phase shift is related to the unit cell length.

To obtain the surface states, we modified the edge cells by adjusting the length of these cells $d_0$ and the width of their central sections $w_0$. This leads to a change in the phase delay and shunt capacitance in the edge unit cells of the microstrip, which change the electrical reactance. Thus, it is possible to match the MSTL impedance to the $Z_0$ input impedance.

We investigated the spatial field distribution for two surface modes of the MSTL structure. The lower and higher frequency modes have symmetric (S) and antisymmetric (A) field distributions with respect to the center of the MSTL. The existence of S and A modes is due to the mirror symmetry of the system. For systems with mirror symmetry all stationary solutions can be grouped into symmetric and antisymmetric modes [7].

We studied the dependence of the width of the surface mode peaks in the transmission spectra on $w_0$. The results of numerical simulation are shown in Fig. 1(b). It can be seen that when the peak doublet is shifted to the edges of the band gap, the peaks width increases faster than their separation. As a result, the peaks that form the doublet merge into one broad peak.

The frequency difference between the A and S modes decreases with the number of MSTL cells (Fig. 1(c)). For a larger number of cells $N$, the amplitudes of both modes decay almost to zero inside the MSTL. As a result, the high-frequency fields on both surfaces do not affect each other. Therefore, the phase difference between them (0 and $\pi$ observed for the surface modes S and A, respectively) is indifferent. Thus the surface states S and A become practically degenerate.

## III. SURFACE STATES: EXPERIMENTAL AND NUMERICAL RESULTS

The results of the simulated transmission spectrum ($S_{21}$ parameter) are included in Fig. 2(a) with brown-dashed and red-dashed lines for the reference MTLS system and the MTLS where the edge cells distorted geometry, respectively. The numerical results reproduce well the features found in the measurements (solid lines), with clear two transmission peaks inside the frequency gap. However, there is a shift of the high frequency edge of the gap, and the shift of the transmission peaks inside the gap toward lower frequencies in simulations. These effects we attribute to the frequency dependence of dielectric constant, which was not included in numerical simulations.

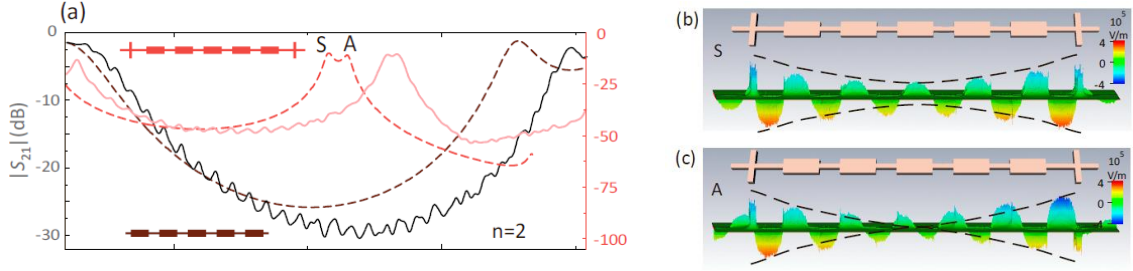

Fig. 2. (a) Transmission spectrum $S_{21}$ for MSTL composed of 5 bulk cells in the absence of edge cells (brown and black line), compared to $S_{21}$ spectrum for the same structure completed with the edge cells (red and pink lines). The double peak in the gap is a signature of the microwave tunneling via surface states. The experimental and numerical data (MEEP) are plotted using solid and dashed lines, respectively. The spatial profiles of: (b) symmetric surface mode (S) and (c) antisymmetric surface mode (A) calculated numerically, which correspond to the double peak visible in (a). The hand-drawn black dashed lines in (b) and (c) show the exponential decay of the amplitude for S and A modes, respectively.

## IV. CONCLUSIONS

We investigated experimentally and numerically the transmission of microwaves through the microstrip transmission line of periodically modulated width. We showed that inside the wide frequency gap, formed in the transmission spectra due to periodicity, the modes with amplitude localized at both terminations of the microstrip allow tunneling transmission through the whole structure.

Also, we demonstrated that the peaks of surface states are split into symmetric and antisymmetric modes, for which the electromagnetic field oscillates on both terminations of the microstrip in-phase and out-of-phase, respectively. Both are suitable for a high transmission of microwaves through the microstripe. We showed that by adjusting values of the geometrical parameters of the edge cells, we can shift the frequency of the surface modes inside the stop band, tune the evanescent coupling between them and tailor the frequency splitting between the symmetric and antisymmetric surface modes.

ACKNOWLEDGEMENT

This work was supported by NCN of the Poland, project OPUS-LAP no 2020/39/I/ST3/02413.